\documentclass[a4paper,fleqn]{cas-sc}

\usepackage[authoryear]{natbib}
\usepackage{float}
\usepackage{aas_macros}
\usepackage{lineno}
\def\tsc#1{\csdef{#1}{\textsc{\lowercase{#1}}\xspace}}
\tsc{WGM}
\tsc{QE}
\begin{document}
\let\WriteBookmarks\relax
\def\floatpagepagefraction{1}
\def\textpagefraction{.001}

\shorttitle{Dynamics of 2021 PH27 - 2025 GN1}    

\shortauthors{Carbognani et al.}  

\title [mode = title]{Investigation of the dynamics and origin of the NEA pair 2021 PH27 and 2025 GN1}  


%\tnotetext[1]{<tnote text>} 


\author[1]{Albino Carbognani}[orcid=0000-0002-0737-7068]

%\cormark[1]


\ead{albino.carbognani@inaf.it}

\ead[url]{https://www.oas.inaf.it/}

\credit{Conceptualization of this study, analysis of NEA pairs and possible fragmentation mechanisms.}

\affiliation[1]{organization={INAF - Osservatorio di Astrofisica e Scienza dello Spazio},
            addressline={Via Gobetti 93/3 }, 
            city={Bologna},
            citysep={}, % Uncomment if no comma needed between city and postcode
            postcode={40129}, 
            country={Italy}}

\author[2, 3]{Marco Fenucci}[orcid=0000-0002-7058-0413]

% Footnote of the second author

% Email id of the second author
\ead{Marco.Fenucci@ext.esa.int}

\ead[url]{https://neo.ssa.esa.int/}

\credit{Study of NEAs orbits}

\affiliation[2]{organization={ESA ESRIN/PDO/NEO Coordination Centre},
            addressline={Largo Galileo Galilei, 1}, 
            city={Frascati (RM)},
            %citysep={}, % Uncomment if no comma needed between city and postcode
            postcode={00044}, 
            country={Italy}}

\affiliation[3]{organization={Deimos Italia s.r.l.},
            addressline={Via Alcide De Gasperi, 24}, 
            city={San Pietro Mosezzo (NO)},
            %citysep={}, % Uncomment if no comma needed between city and postcode
            postcode={28060}, 
            country={Italy}}

\author[4,5]{Toni Santana-Ros}[orcid=0000-0002-0143-9440]

\affiliation[4]{
    organization={Departamento de Física, Ingeniería de Sistemas y Teoría de la Señal, Universidad de Alicante},
    addressline={Carr. San Vicente del Raspeig, s/n},
    city={San Vicente del Raspeig, Alicante},
    %citysep={},
    postcode={03690},
    country={Spain}
}

\affiliation[5]{
    organization={Institut de Ciències del Cosmos (ICCUB), Universitat de Barcelona},
    addressline={ c. Martí i Franquès, 1},
    city={Barcelona},
    %citysep={},
    postcode={08028},
    country={Spain}
}

\author[6]{Clara E.~Mart\'inez-V\'azquez}[orcid=0000-0002-9144-7726]
\affiliation[6]{
%organization={International Gemini Observatory/NSF NOIRLab},
organization={NSF NOIRLab},
addressline={ 670 N. A'ohoku Place},
city={Hilo, Hawai'i,},
postcode={96720},
country={USA}
}

\author[2]{Marco Micheli}[orcid=0000-0001-7895-8209]

% Footnote of the thirt author

% Email id of the third author
\ead{Marco.Micheli@esa.int}

% URL of the third author
\ead[url]{https://neo.ssa.esa.int/}

\credit{Asteroids physical characterization}

\cortext[1]{Corresponding author}

\fntext[1]{}



\begin{abstract}
We analyse the association between the NEAs 2021~PH27 and 2025~GN1, which share similar heliocentric Keplerian elements and the same taxonomic classification. First, we confirm the spectral similarity by getting independent colours measurements of 2025~GN1 and confirming that they are both X-type. From numerical integration of the orbits up to 100 kyr in the past, taking into account relativistic corrections, we found that the two asteroids experienced five similar flybys with Venus, but none of them were closer than the Roche limit. The perihelion distance also reached values between 0.1 and 0.08 au about 17/21 kyr and 45/48 kyr ago, but still well outside the Roche limit with the Sun. So, the origin of the pair by tidal disruption of a progenitor rubble-pile asteroid appears unlikely. On the other hand, we found periods lasting several thousand years where the perihelion was below 0.1 au, and this can lead to thermal fracturing of the surface. We found that the rotation period of the primary and the mass ratio secondary/primary make the pair indistinguishable from the binary systems known among the NEAs, and the YORP effect can double the rotation period of 2021 PH27 in $150 \pm 50$ kyr. So it is plausible that the pair was formed by the rotational disintegration of a rubble-pile asteroid due to anisotropic gas emission or the YORP effect, which formed a binary system that later dissolved due to the internal dynamics of the pair. We are unable to give a value for the separation age; we can only say that it occurred more than 10.5 kyr ago and may have occurred between 17/21 kyr ago during the last and longer phase of lower perihelion distance. In this scenario, little meteoroids released in space due to the fragmentation event are still near the pair's orbit and can generate a meteor shower in Venus' atmosphere. 
\end{abstract}

\begin{highlights}
\item Dynamic evolution of 2021~PH27 and 2025~GN1.
\item Origin of the pair 2021~PH27 and 2025~GN1. 
\item 2021~PH27 and 2025~GN1 appear as an ex-binary system.
\end{highlights}

% Keywords
% Each keyword is separated by \sep

\maketitle

%%%%%%%%%%%%%%%%%%%%%%%%%%%%%%%%%%%%%%%%%%%%%%%%%%

%%%%%%%%%%%%%%%%% BODY OF PAPER %%%%%%%%%%%%%%%%%%

\section{Introduction}
\label{sec:intro} 
\noindent In a previous paper, we explored the possibility that the near-Earth asteroid (NEA) 2021~PH27 is also an active asteroid like (3200) Phaethon, the progenitor of the Geminid meteor shower. This idea comes from the fact that the perihelion of 2021~PH27 is lower than Phaethon’s perihelion, and the same physical mechanisms that make Phaethon active could also operate on 2021~PH27 \citep{Carbognani2022}. However, the minimum orbit intersection distance (MOID) of 2021~PH27 with respect to Earth is large, about $0.22$ au, while in the case of Phaethon, it is only $0.0188$ au. Instead, Venus' MOID is low with a value of about $0.0146$ au, and 2021~PH27 could be the Phaethon of Venus. Unfortunately, the bright fireballs in Venus' atmosphere are very difficult to observe, and this hypothesis remains unconfirmed.\\
The recent discovery of the NEA 2025~GN1 in a heliocentric orbit close to that of 2021~PH27 \citep{Sheppard2025} reopened the question. These two asteroids share similar orbits, so it is possible that there could be other smaller meteoroids on the same orbit. Hence, there is a need to understand the dynamic origin of this pair of asteroids.

\subsection{Asteroid pairs}
\noindent Asteroid pairs consist of two asteroids with very similar orbital elements but not bound together: they are found between the main-belt, Hungaria, and NEAs populations. The discovery of asteroid pairs dates back to 2008, when \citet{Vokro2008} announced the existence of about 60 pairs among main-belt asteroids. Around 25\% of the pairs found were associated with recently formed asteroid families. Still, the presence of 45 pairs with orbital elements so similar as to make a random coincidence very unlikely (less than 1\%) remained to be explained. The possible mechanisms proposed by \citet{Vokro2008} to explain the existence of asteroid pairs are: (1) catastrophic collision, (2) rotational fission due to the Yarkovsky-O'Keefe-Radzievski-Paddack (YORP) effect, or (3) dissociation of binary systems. Note that the YORP effect can change the rotation period of an asteroid quickly compared to the asteroid's lifetime if the size is less than about 10 km, so mechanism number two is operative for all NEAs.\\ 
Photometric observations by \citet{Pravec2010} on 35 pairs of main-belt asteroids establish that the pair primary (i.e, the largest component of the two) rotates quickly, near critical fission frequency, and that the secondary/primary mass ratio is not greater than $\mu\approx 0.2$. This indicates that the dominant formation mechanism is (2) because, from the rotational fission model, we know that binary systems with $\mu < 0.2$ have positive free energy (defined as the total energy minus the self-potentials of each component) and can escape each other. In contrast, for $\mu > 0.2$, the free energy is negative, and the pair is bound to form a real binary system, i.e., the energy is insufficient to disrupt it \citep{Scheeres2007, Pravec2010}. \\
Therefore, in most cases, the parent of a pair of main-belt asteroids is a body with a rubble pile structure, which, due to the YORP effect, decreases its rotation period until it exceeds the value of its cohesionless spin barrier, equal to \citep{Pravec2000, Carbognani2017}:

\begin{equation}
P_{lim} = \frac{3.301}{\sqrt\rho}.
\label{eq:limit_period}
\end{equation}

\noindent In Eq.~\eqref{eq:limit_period}, the period is in hours, and $\rho$ is the average density of the asteroid in $\textrm{g}/\textrm{cm}^3$. This causes the body to split into at least two components due to weak gravity, and if the secondary component is sufficiently small, it can acquire enough angular momentum from the primary's fast rotation to separate. In practice, pairs of asteroids are binary systems formed due to YORP, which break apart soon after their formation.

\begin{table*}
	\centering
	\caption{Physical properties of the known NEA pairs in discovery order from top to bottom. Note that for Phaethon, there is a triplet. The $P$ value is the rotation period with uncertainty; a question mark indicates that it is unknown. Tax is the taxonomic class, MOID is the Minimum Orbit Intersection Distance with the Earth, $q$ is the perihelion distance, $i$ is the orbital inclination, $H$ is the absolute magnitude, $D$ is the estimated diameter, $\mu=m_2/m_1$ is the secondary/primary mass ratio (see Eq.~\eqref{eq:mass_ratio}), while $D_{SH}$ and $D_{D}$ are the orbital dissimilarity functions of \cite{Southworth1963} and \cite{Drumond1981}. Physical data are from the Near-Earth Objects Coordination Centre (\url{https://neo.ssa.esa.int/home}) or the bibliography cited in the text. The last line also reports the pair analyzed in the paper.}
	\label{tab:NEA_physical_data}
        \setlength\tabcolsep{2pt} % default value: 6pt
	\begin{tabular}{lcccccccccc} 
		\hline
	Designation & $P$ (h)  & Tax & MOID (au) & $q$ (au) & $i$ (deg) & $H$ & D (m)  & $\mu$ & $D_{SH}$ & $D_{D}$\\
		\hline
(3200) Phaethon   &  $3.603 \pm 0.006$ &   B     &  0.0195  & 0.14 & 22.3 & 14.4  & 5100  &       &        & \\
(155140) 2005 UD  &  $5.249 \pm 0.004$ &   B     &  0.0772  & 0.16 & 28.6 & 17.5  & 1300  & 0.014 & 0.731  & 0.790  \\
(225416) 1999 YC  &  $4.495 \pm 0.001$ &   ?     &  0.2475  & 0.24 & 38.3 & 17.3  & 1650  & 0.016 & 1.009  & 0.944 \\
\hline
(1566) Icarus     &  $2.2736 \pm 0.00005$ &   Q     &  0.0338  & 0.19 & 22.8 & 16.6  & 1440  &       &         &\\
2007 MK6          &   ?                            &   ?     &  0.0872  & 0.19 & 25.1 & 20.3  & 300   & 0.006 & 0.058   & 0.031 \\
\hline
2017 SN16         &   ?         &   V     &  0.0938  & 0.87 & 13.4 & 23.3  & 80  &         &         & \\
2018 RY7          &   ?         &   V     &  0.0947  & 0.87 & 13.3 & 24.4  & 50  & 0.218   &  0.002  & 0.005 \\
\hline
2015 EE7          &   $9.420 \pm 2$ &   Sq    &  0.0653  & 1.00 & 27.3 & 20.2  & 260  &        &         & \\
2015 FP124        &   ?                      &   Q     &  0.0654  & 1.00 & 27.4 & 22.2  & 130  & 0.063  &  0.002  & 0.002 \\
\hline
2019 PR2          &   ?         &   D     &  0.2323  & 1.16 & 10.7 & 18.7  & 700  &        &          &  \\
2019 QR6          &   ?         &   D     &  0.2322  & 1.17 & 10.9 & 20.0  & 400  & 0.166  &  0.0003  & 0.0001 \\
\hline
(33342) 1998 WT24 &   $3.697 \pm 0.002$ &   E     &  0.0100  & 0.42 & 7.4  & 18.0  & 400 &         &       &  \\
(745311) 2010 XC15 &  $2.673 \pm 1$     &   E     &  0.0018  & 0.42 & 8.2  & 21.5  & 100 & 0.008   & 0.041 & 0.016 \\
\hline
\textbf{2021 PH27}         &   $3.49^{*}$ &   X     & 0.225    & 0.13 & 31.9 & 17.7  & 1200 &        &        & \\
\textbf{2025 GN1}          &   ?          &   X     & 0.224    & 0.14 & 32.8 & 20.0  & 400  & 0.037  & 0.027  & 0.015 \\  
		\hline
        \multicolumn{6}{l}{$^{*}$ Unpublished value.}\\ 
	\end{tabular}
\end{table*}

\subsection{The NEA pairs}
\noindent The same splitting mechanisms operating in the main-belt are also active among the dynamic population of NEAs, so we can expect pairs of asteroids also here, although in fewer numbers, mainly because the orbits of the NEAs are chaotic, with time scales lower than those of the NEA lifetime. A list of physical properties of well-known NEA pairs is given in Table~\ref{tab:NEA_physical_data}, where we also report the orbital inclination $i$ to highlight how for pairs of NEAs, the values range between about 10 and 40 degrees. There are no known pairs with much lower inclinations, probably because the higher the inclination, the fewer gravitational perturbations from the planets in the inner Solar System, and this extends the pair's lifetime.\\
The first NEAs pair to be discovered was (3200) Phaethon and (155140) 2005~UD in 2006 \citep{Ohtsuka2006}. Phaethon was already known as the progenitor body of the Geminids meteor shower, and a search was made for any large fragments of what was believed to be an extinct comet. By numerically integrating the position of these two asteroids backwards and forward in time, they demonstrate that the evolution of the orbital elements was the same, only shifted in time by approximately 4600 years, a sign of a common origin. The Daytime Sextantids meteor shower is associated with asteroid 2005~UD \citep{YeQuanZhi2018}, which is likely a fragment of Phaethon himself, and the Daytime Sextantids are a part of the Phaethon-Geminid Stream Complex \citep{Kipreos2022}. The possible age of separation for this pair is about 100 kyr ago or even before, and this implies that the separation event is older than the Geminids ejection in space \citep{Hanus2016}. In 2008, this pair was also joined by 1999~YC \citep{Ohtsuka2008}. Due to the larger semi-major axis compared to the original pair, it is hypothesised that 1999~YC had a close fly-by with a terrestrial planet, which led to a wider orbit. \\
The second NEA pair, (1566) Icarus - 2007~MK6, was discovered in 2007 using the same time-lag technique previously employed for Phaethon and has an estimated separation age of approximately 1 kyr ago \citep{Ohtsuka2007}. In this discovery paper, there is mention of a possible daytime faint meteor shower associated with the pair; it is the Taurid-Perseid shower, discovered with radar in 1973 and with a maximum rate on June 18th \citep{Sekanina1973}. This is the first hint of an association between a pair of NEA asteroids and a meteor shower, but Icarus was already suspected of being associated with the meteor shower of the Daytime Arietids \citep{Duncan1988}.\\
In 2019 \citet{delaFuente2019} identified the pair of NEA 2017~SN16-2018~RY7. These two asteroids do not show any increase in their mean relative longitude because they are in a 3:5 mean motion resonance with Venus. From the dynamic simulations, it appears that the pair cannot have been trapped in this resonance by chance; therefore, the two most probable hypotheses for its formation are a split of a rubble pile asteroid due to the YORP effect or the disintegration of a binary system due to the resonance.\\
The NEA 2017~SN16-2018~RY7 pair, together with 2015~EE7 - 2015~FP124, was the subject of a paper by \citet{Moskovitz2019}. This work presents the results of spectroscopic observations, which show how the pair members belong to the same taxonomic class. The fact that the pair 2017~SN16-2018~RY7 members are both V-type asteroids, a fairly rare taxonomic type among the NEAs (2-4\%), favours the pair reality. Furthermore, dynamic simulations show that the two pairs were formed by dissociating a parent binary system. The age of separation was determined only for 2017~SN16 - 2018~RY7 (less than 10 kyr), while for 2015~EE7-2015~FP124, a value is not given due to the uncertainty of the orbits. \\
In \citet{Fatka2022}, the pair of NEA~2019 PR2 - 2019~QR6 is studied, finding that they have a common origin because (1) they are of the same spectral type D (scarce among NEAs) and (2) the orbital evolution in the past suggests a separation event that occurred approximately 300 yrs ago. This very recent age for separation makes this the youngest known NEA pair. To explain today's orbits, it was necessary to hypothesise the presence of cometary activity after the separation, but now the pair members show no current signs of cometary activity. The latest NEA pair identified is 1998~WT24 $-$ 2010~XC15 \citep{Beniyama2023}. These asteroids are rare E-type, and the motion integrals are very similar, indicative of a common origin. \\
The most interesting thing about the pairs of NEAs shown in Table~\ref{tab:NEA_physical_data} is the column, where the mass ratio is reported: $\mu=m_2/m_1$, where $m_2$ and $m_1$ denote the masses of each pair's smaller and larger members. As in \citet{Vokro2008}, we estimated $\mu$ from the absolute magnitude $H_1$ and $H_2$ of the two pairs' members and assuming the same albedo and mean density for each asteroid of the pair, so:

\begin{equation}
\mu=10^{-0.6(H_2-H_1)}.
\label{eq:mass_ratio}
\end{equation}

\begin{figure}
    \centering
    \includegraphics[width=1.0\textwidth]{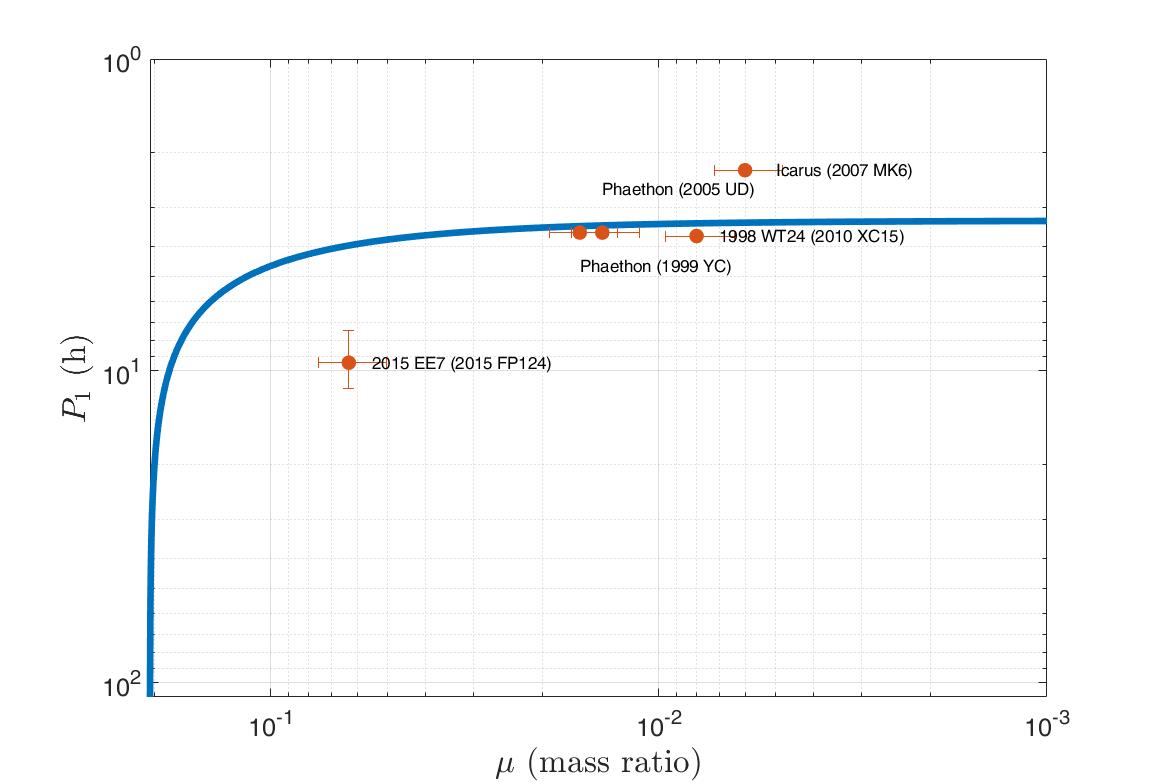}
    \caption{Plot of the primary rotation periods (when known) as a function of the mass ratio for the NEA pairs listed in Table~\ref{tab:NEA_physical_data}. The uncertainty of the mass ratio is computed from Eq.~\eqref{eq:mass_ratio} assuming a 0.1 mag uncertainty over the absolute magnitude given by the Minor Planet Center. The continuous curve, which is purely indicative and is not a best fit, is an example of the predictions of the post-separation model fission from \citet{Pravec2010} with asteroid's parameters shape $a_1=1$, $b_1=0.8$, $c_1=0.2 b_1$; mean density $\rho = 1000~\text{kg}~\text{m}^{-3}$ and starting orbit semimajor-axis of the satellite $A_{ini}=2$. The rotation period value of 2015 EE7 is uncertain. }
    \label{fig:NEA_pairs_plot}
\end{figure}

\noindent The $\mu$ value is never larger than about 0.2, as expected for a binary separation origin \citep{Pravec2010}, see Fig.~\ref{fig:NEA_pairs_plot}. In the case of low values of $\mu$, where the rotation period of the primary is also known, this value appears close to the value of the spin barrier. Unfortunately, the rotation periods of the pairs with high values of $\mu$, 2017~SN16 $-$ 2018~RY7, is not know: it would have been interesting to check if the period of the primary is consequently longer. No close passes of the primary of this pair are expected until at least October 10, 2058, when it will be 1.5 million km from Earth at magnitude +20.2.

\noindent In this work, we present evidence that the candidate NEA pair 2021~PH27 - 2025~GN1 has a common physical origin and that the parent asteroid has probably undergone a rotational disintegration, forming an unstable binary system that has undergone separation, creating the actual pair. The paper is organised as follows: in Section~\ref{sec:2021PH27}, we present the physical characteristics of 2021~PH27 and 2025~GN1, while in Section~\ref{sec:age_separation}, we examine the common dynamical evolution of the two asteroids. In Section~\ref{sec:discussion}, we discuss the results obtained, and finally, we provide our conclusions.

\section{Physical characteristic of 2021~PH27 and 2025~GN1}
\label{sec:2021PH27}
\begin{figure}
    \centering
    \includegraphics[width=0.5\linewidth]{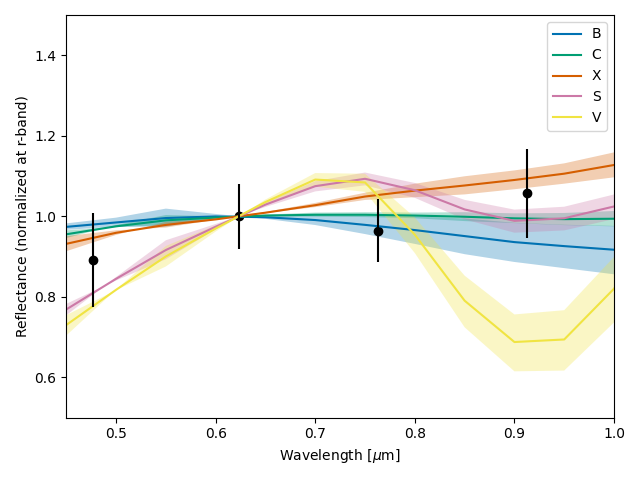}
    \caption{Colours of 2025~GN1 obtained from observations of the Gemini South 8.1 m telescope on 16 April 2025. Template spectra for B-, C-, X-, S-, and V-type asteroids are also reported for comparison, and for each reflectance spectrum, the variation within the taxonomic class is indicated \citep{DeMeo2013}.}
    \label{fig:spectral-comp}
\end{figure}

The NEA 2021~PH27 was discovered during the twilight of Aug 13, 2021, using the V\'ictor M. Blanco Telescope of 4 metres aperture of the Cerro Tololo Observatory \citep{Sheppard2022}. The discovery was reported in the electronic circular MPEC 2021~Q41\footnote{\url{https://www.minorplanetcenter.net/mpec/K21/K21Q41.html}} published by the Minor Planet Center (MPC) on Aug 21, 2021. The NEA 2025~GN1 was discovered on Apr 4, 2025, with the same instrument and announced in the MPEC 2025-G124\footnote{\url{https://www.minorplanetcenter.net/mpec/K25/K25GC4.html}}. \\
The similarity of the osculating orbital elements of 2021~PH27 and 2025~GN1, see  Table~\ref{tab:pair_orbital_elements}, has drawn attention to these two asteroids. We now know that these two NEAs also share not only their orbits but also the same colour indices. From \cite{Sheppard2025} the colour indices for 2021~PH27 are $g^{'}-r^{'}=0.58 \pm 0.02$, $r^{'}-i^{'}=0.12 \pm 0.02$ while for 2025~GN1 $g^{'}-r^{'}=0.55 \pm 0.06$, $r^{'}-i^{'}=0.14 \pm 0.04$. These values are compatible with each other within the uncertainty, supporting the interpretation that the two objects form a dynamically related asteroid pair of common origin. %, and the two NEAs are classified as X-type asteroids. 
%
% Toni's paragraph
To independently confirm the results on the similarity in physical characteristics, we obtained photometric observations of asteroid 2025~GN1 with the Gemini South 8.1 m telescope using the GMOS-S instrument on Apr 16, 2025. Two groups of eight exposures were acquired in each of the $g^{'}, r^{'}, i^{'}$, and $z^{'}$ filters and combined into single stacked images per group to improve the signal-to-noise ratio. From these stacks, we measured median magnitudes of $g^{'} = 21.91 \pm 0.12$, $r^{'} = 21.34 \pm 0.06$, $i^{'} = 21.28 \pm 0.06$, and $z^{'} = 21.14 \pm 0.08$, yielding colours of $g^{'}-r^{'} = 0.58 \pm 0.13$, $r^{'}-i^{'} = 0.06 \pm 0.09$, and $i^{'}-z^{'} = 0.14 \pm 0.10$. These values are consistent with those reported by \citet{Sheppard2025}, and place 2025~GN1 within the neutral, slightly red region of SDSS colour space, most consistent with an X-type asteroid, see Fig.~\ref{fig:spectral-comp}. %2025 GN1 shares nearly identical orbital elements with 2021 PH27, which exhibits similar colors (g′–r′ = 0.58 ± 0.02, r′–i′ = 0.12 ± 0.02, i′–z′ = –0.08 ± 0.05; Sheppard et al. 2025), supporting the interpretation that the two objects form a dynamically related asteroid pair of common origin.

Regarding the size, we can estimate it using the equation that connects the effective diameter with the absolute magnitude and the geometrical albedo \citep{bowell-etal_1989, Harris2002}:
\begin{equation}
    D=\frac{1329}{\sqrt{p_V}} 10^{-H_V/5}.
    \label{eq:diameter}
\end{equation}

\noindent In Eq.~\eqref{eq:diameter}, the diameter $D$ is in km, $H_V$ is the absolute magnitude in $V$ band, while $p_V$ is the geometric albedo in the same band. For X-type asteroids, the mean geometric albedo is $p_V=0.098 \pm 0.081$ \citep{Usui2013}. Assuming the absolute magnitude given by the MPC with $H_V=17.67 \pm 0.26 $ for 2021~PH27 and $H_V=20.06 \pm 0.36$ for 2025~GN1, we get a diameter of $1.2 \pm 0.5$ km for the first asteroid, and a value of $0.4 \pm 0.2$ km for the second. As the uncertainty value for the absolute magnitude, we took the RMS of the fit with the Bowell HG model. The data from MPC contain astrometric observations made by different surveys with different photometric bands and different levels of photometric calibration \citep[see][]{Hoffmann2024}, so the values of absolute magnitude, and consequently the diameter values, should be considered with caution, although we expect the order of magnitude to be as indicated. Furthermore, in the MPC data, there is no explicit value for absolute magnitude uncertainty. \\

\begin{figure}
    \centering
    \includegraphics[width=0.98\linewidth]{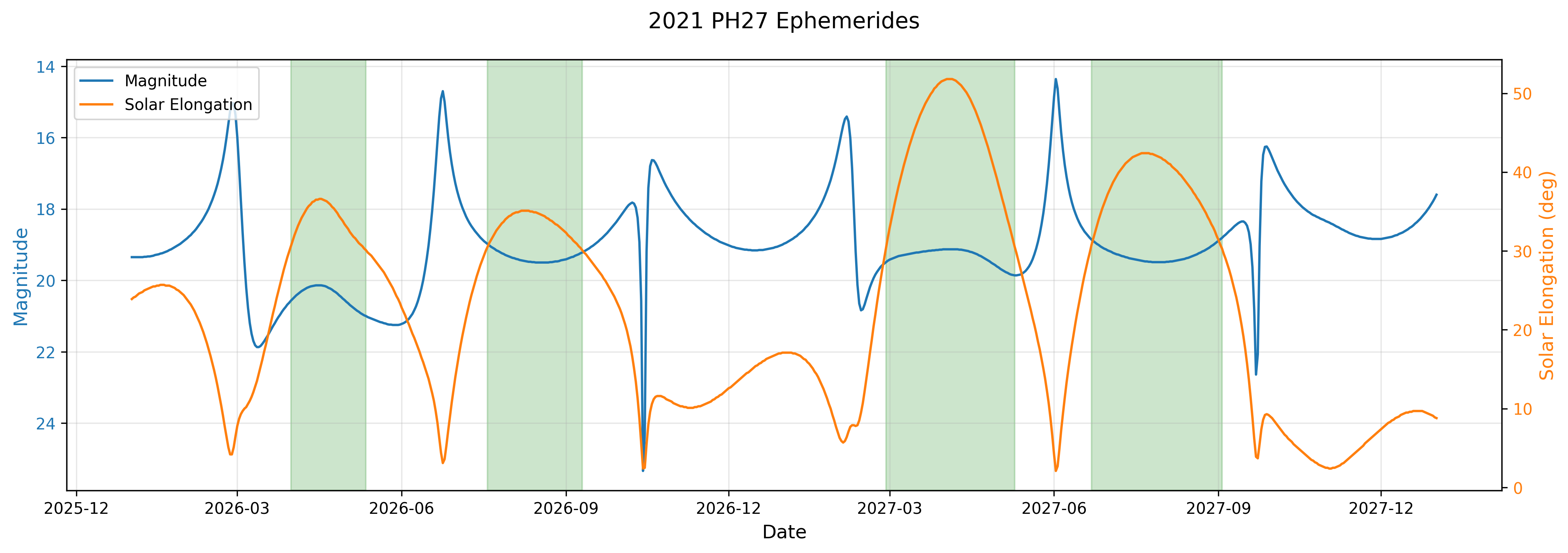}
    \includegraphics[width=0.98\linewidth]{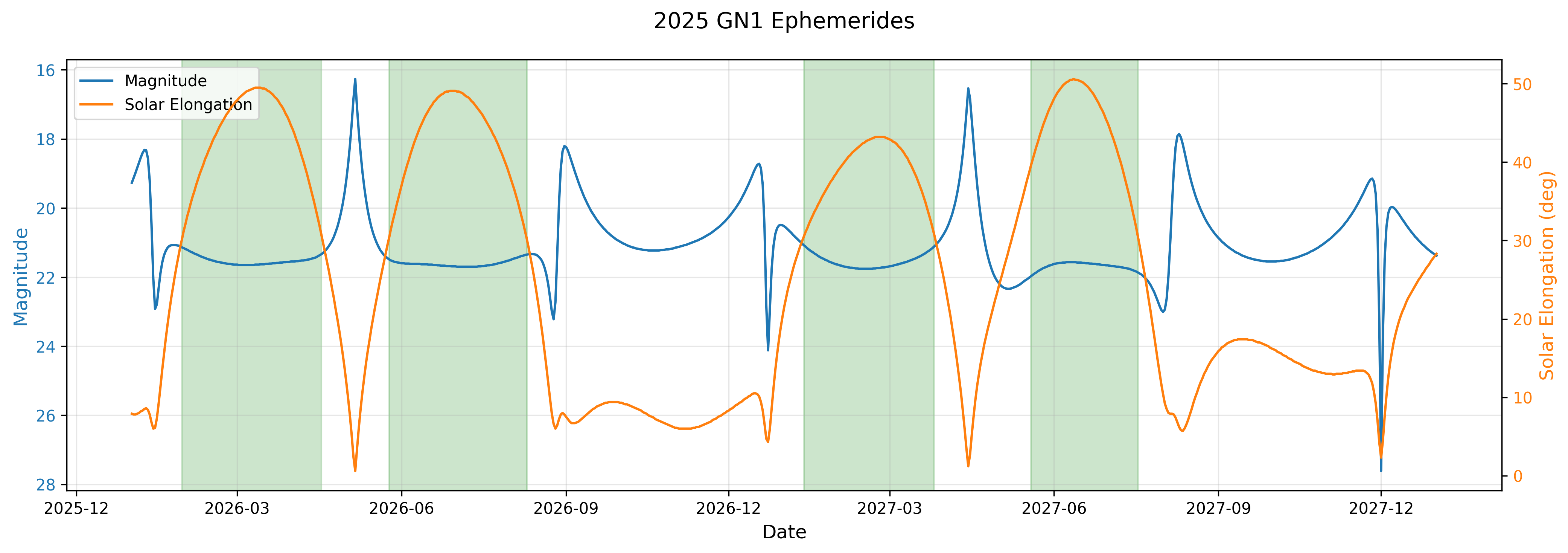}
    \caption{Visual magnitude (blue curve) and solar elongation (orange curve) of 2021~PH27 (top panel) and 2025~GN1 (bottom panel) until 1 January 2028. The green shaded areas correspond to visibility windows, identified by visual magnitude smaller than 22 and solar elongation larger than 30 deg in absolute value.}
    \label{fig:ephem2years}
\end{figure}

About the rotation period of 2021~PH27, there is an unpublished lightcurve data by T. Santana-Ros et al., that suggests a period of $3.49 \pm 0.01$ h\footnote{\url{https://archive.gemini.edu/programinfo/GS-2024A-DD-101}} based on visible measurements made in Feb, Mar, and Apr 2022 \citep{Muller2022}. The lightcurve will become public in the near future. No information is known regarding the rotation period of 2025~GN1. In \cite{Sheppard2025}, the observation time of the two asteroids, less than an hour, was too short to allow the determination of a rotation period, even if a magnitude variation of 0.1 mag was noted for 2021~PH27 and 0.2 mag for 2025~GN1. Figure~\ref{fig:ephem2years} show the absolute magnitude and the solar elongation of the two asteroids until 1 January 2028. The next favourable observation windows for 2025~GN1 are in mid-March 2026 and June 2026, when the asteroid will reach a visual magnitude of about 21.8 at a solar elongation of about 50 deg. While still faint, photometric measurements to extrapolate a lightcurve may be possible to perform with large-aperture telescopes. The next observability windows for 2021~PH27 are around April 2026 and mid-August 2026, although at a maximum elongation of about 35 deg. The asteroid will reach a larger elongation of about 50 deg only in April 2027.  
\\

In Table~\ref{tab:NEA_physical_data}, the columns of the orbital dissimilarity functions show that the 2021~PH27-2025~GN1 candidate pair has values intermediate between Phaethon's pair and the 2015~EE7 pair, so it is not distinguishable from well-studied pairs. To close the section about the physical characterization of 2021~PH27 and 2025~GN1, we estimated how likely it is to randomly find a pair like this among the general population of NEAs. To this purpose, we used the classical \cite{Southworth1963} orbit similarity criterion and the synthetic population of 802,000 NEAs by \citet{Granvik2018}. The $D_{SH}$ value for the pair 2021~PH27 and 2025~GN1 is 0.027, which is slightly lower than the value for the Geminids and Phaethon with $D_{SH} \approx 0.031$. By randomly sampling 1 billion possible pairs from the synthetic population, we estimated a probability of $4\cdot 10^{-8}$ of having some random NEA pair with $D_{SH} < 0.027$. Considering that X-type asteroids among NEAs are about 0.19 of the total \citep{Hromakina2021}, we can estimate an overall probability of $10^{-9}$ of having a pair of NEAs with $D < 0.027$, both of X-type. This is low enough to consider a random pair, such as 2021~PH27-2025~GN1, unlikely.

\section{Past dynamical evolution}
\label{sec:age_separation}
The past dynamical evolution of the pair 2021~PH27 and 2025~GN1 was investigated by backwards numerical simulations. Numerical integrations were performed with the \texttt{mercury} software package \citep{chambers-migliorini_1997}, by using a symplectic hybrid integrator able to solve close approaches with planets \citep{chambers_1999}. The dynamical model included the gravitational attraction of the Sun and of all the planets from Mercury to Neptune. Masses and initial positions of the planets at epoch 60800 MJD were retrieved from the JPL Ephemeris DE441 \citep{park-etal_2021}. \\
Due to the low perihelion distance of about 0.13 au, which the two asteroids currently achieve, the effects from general relativity also need to be included in the model. To this purpose, we implemented the potential $U = 3(GM_\odot)^2/(cr)^2$ described in \citet{nobili-roxburgh_1986, nobili-etal_1989}, where $G$ is the universal gravitational constant, $M_\odot$ is the mass of the Sun, $c$ is the speed of light, and $r$ is the distance of the body from the Sun. The acceleration associated with this potential exactly reproduces the perihelion precession due to general relativity, which is the main secular effect. This effect has been added to the modified \texttt{mercury} version by \citet{fenucci-novakovic_2022}, which is freely available at GitHub\footnote{\url{https://github.com/Fenu24/mercury}}.

{\renewcommand{\arraystretch}{1.2} % Change row height for this table only
\begin{table*}[!ht]
    \centering
    \caption{Keplerian orbital elements of 2021~PH27 and 2025~GN1 taken from the ESA NEOCC. Errors refer to the formal 1$\sigma$ marginal uncertainties.}
    \begin{tabular}{lrr}
         \hline
         \hline
      Parameter                    & 2021~PH27 & 2025~GN1  \\
         \hline
      Epoch                                  & 60800.0 MJD             & 60800.0 MJD \\ 
      Semi-major axis, $a$ (au)              &  $   0.461756878~ (\pm 1.87 \times 10^{-8})$ &  $   0.46197144~ (\pm 4.25\times 10^{-7})$ \\ 
      Eccentricity, $e$                      &  $   0.711563909~ (\pm 5.79 \times 10^{-7})$ &  $   0.70507949~ (\pm 1.22\times 10^{-6})$ \\ 
      Inclination, $i$ (deg)                 &  $  31.941369202~ (\pm 9.35 \times 10^{-5})$ &  $  32.83622289~ (\pm 1.09\times 10^{-4})$ \\ 
      Longitude of node, $\Omega$ (deg)      &  $  39.396585975~ (\pm 5.79 \times 10^{-5})$ &  $  41.01599281~ (\pm 1.04\times 10^{-4})$ \\ 
      Argument of perihelion, $\omega$ (deg) &  $   8.579583738~ (\pm 3.07 \times 10^{-5})$ &  $   6.10495641~ (\pm 6.44\times 10^{-5})$ \\ 
      Mean anomaly, $\ell$  (deg)            &  $ 140.256704482~ (\pm 1.97 \times 10^{-4})$ &  $ 291.35705002~ (\pm 2.38\times 10^{-4})$ \\ 
      %Normalized RMS                         &         0.598         &         0.575         \\
         \hline
    \end{tabular}
    \label{tab:pair_orbital_elements}
\end{table*}
}

\begin{figure}
    \centering
    \includegraphics[width=0.70\textwidth]{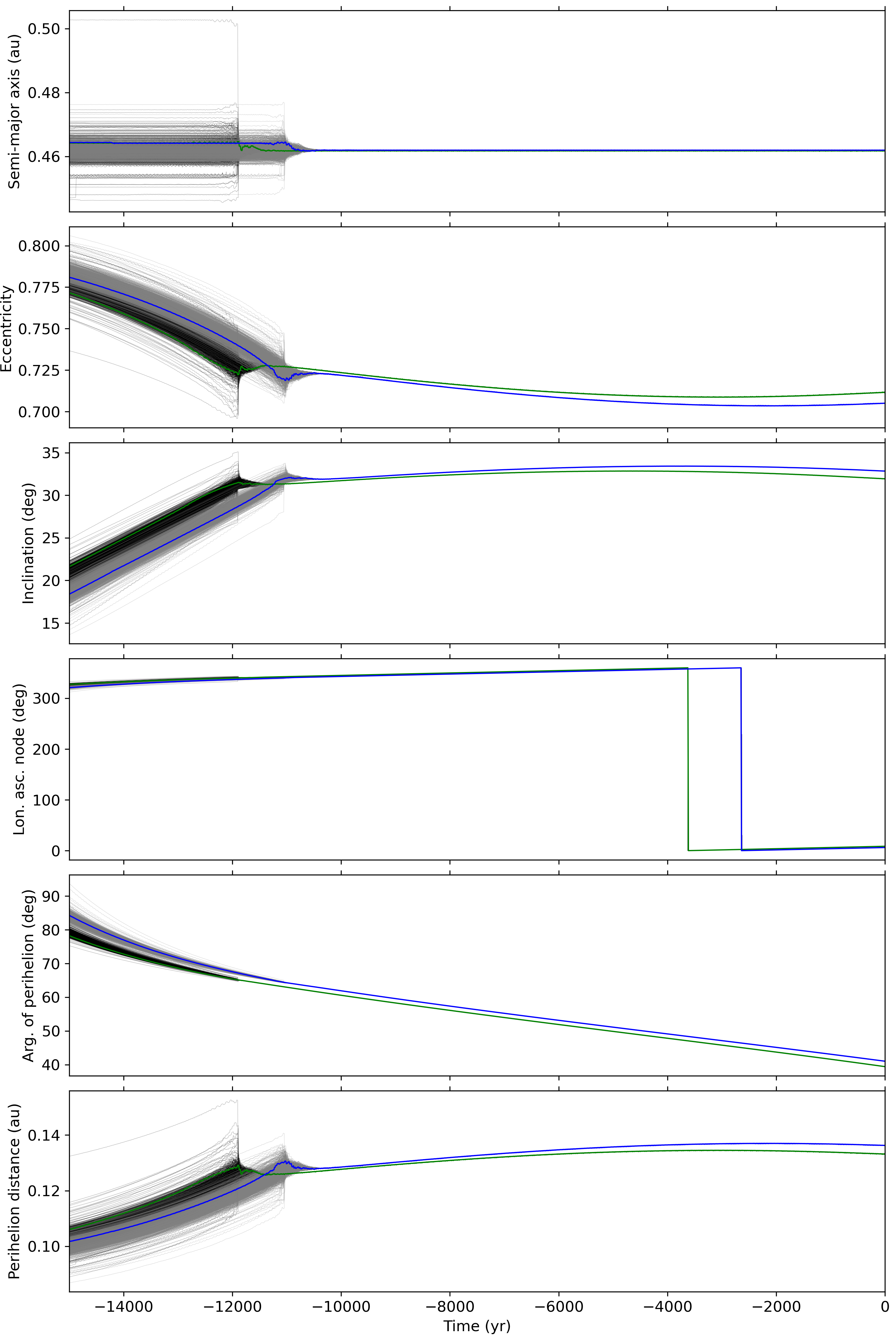}
    \caption{Orbital elements of clones of 2021~PH27 (black), with nominal orbit in green, and of 2025~GN1 (grey), with nominal orbit in blue. }
    \label{fig:clo_elements}
\end{figure}

\begin{figure}
    \centering
    \includegraphics[width=0.48\textwidth]{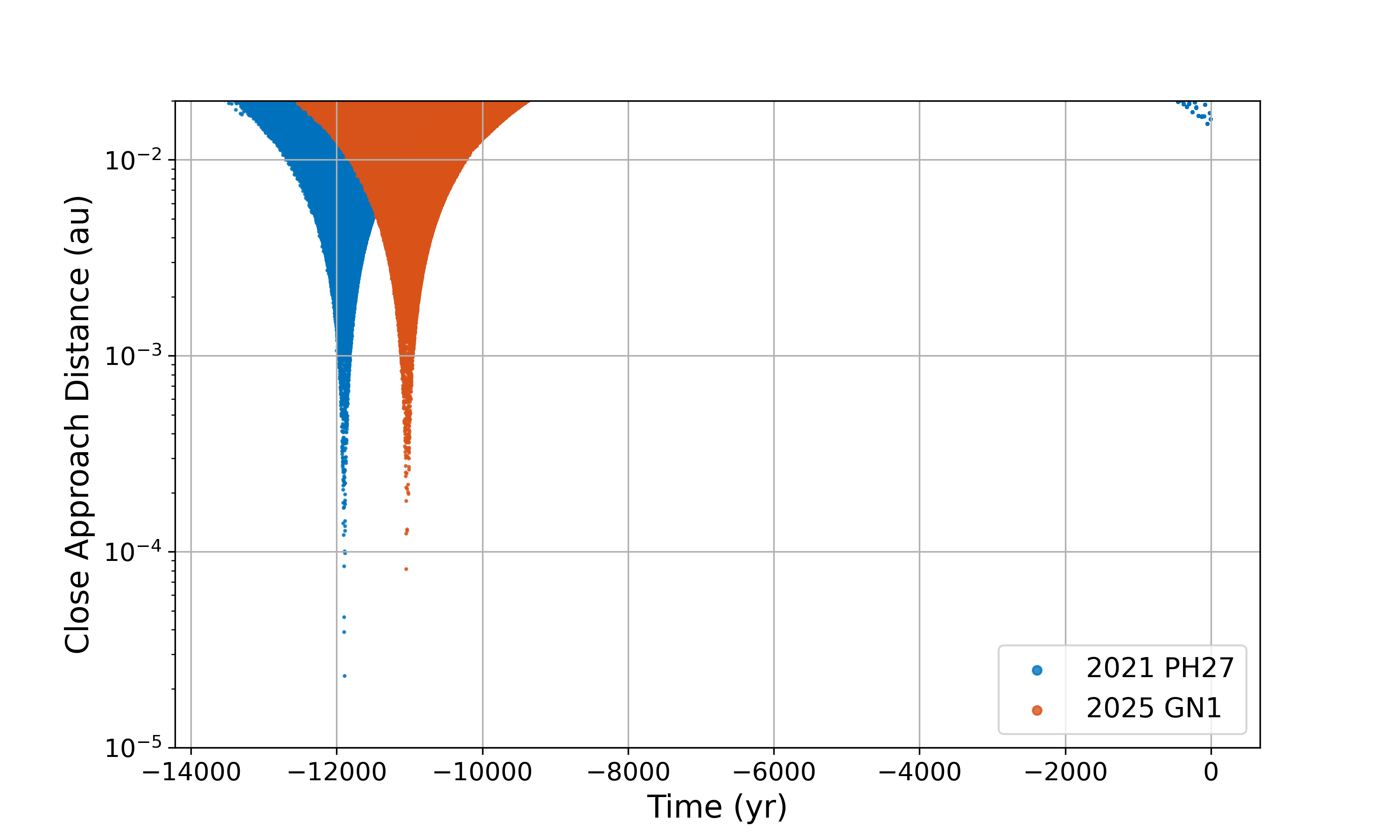}
    \includegraphics[width=0.48\textwidth]{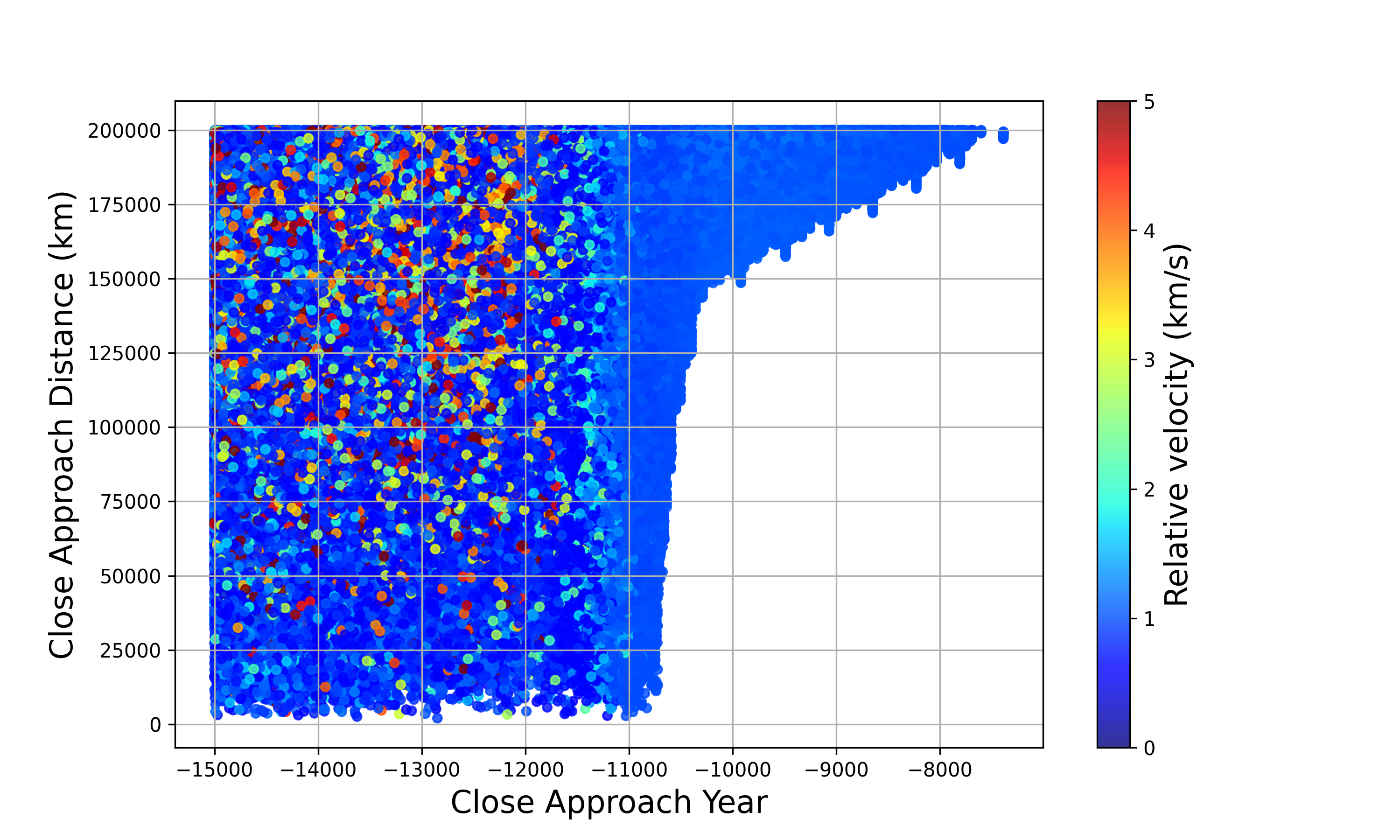}
    \caption{On the left panel, distance of the recorded close encounters with Venus, as a function of time. Blue dots are close encounters registered for 2021~PH27, while the red dots are those for 2025~GN1. On the right panel, a scatter plot of close encounters between 2021~PH27 and 2025~GN1, in the plane of close approach year and close approach distance. The colour is proportional to the relative velocity of the close encounter.}
    \label{fig:Venus_close_encounters}
\end{figure}

For each asteroid, we propagated 1000 orbital clones, which were sampled from the covariance matrix of the orbit by assuming a Gaussian statistics \citep[see also][for details]{fenucci-novakovic_2021}. Nominal orbital elements from the ESA NEO Coordination Centre\footnote{\url{https://neo.ssa.esa.int/}} (NEOCC), together with the formal 1$\sigma$ marginal uncertainties, are reported in Table~\ref{tab:pair_orbital_elements}. The full orbit with the covariance matrix needed to reproduce the results presented here can be directly downloaded from the NEOCC website.\\
Orbits were propagated backwards for 15 kyr with a timestep of 0.5 days, and output coordinates were recorded with a 1-day step. 
The propagation of orbital clones allows also the estimation of the Lyapunov time of the orbit, which gives the timescale for the predictability horizon \citep{Froeschle1984}. After roughly a Lyapunov time, the orbit evolution is dominated by stochastic processes, and the results should be treated statistically. The computed Lyapunov time for 2021~PH27 and 2025~GN1 are 6800 yr and 5900 yr, respectively.
Fig.~\ref{fig:clo_elements} shows the evolution of semi-major axis, eccentricity, inclination, longitude of the ascending node, argument of perihelion, and perihelion distance of the simulated clones of 2021~PH27 and 2025~GN1. The orbital elements do not undergo large changes until $-11$ kyr and $-12$ kyr for 2025~GN1 and 2021~PH27, respectively. 
Note that this timespan is longer than the computed Lyapunov time, however this is not inconsistent with the results obtained. In fact, the Lyapunov time takes into account also the uncertainties in the position of the asteroid, which is dominated by the uncertainties in the mean anomaly \citep{milani-etal_2005b}, while the other elements may have a more regular evolution over a longer period of time \citep{gronchi-milani_2001,fenucci-etal_2022}.
At the two epochs mentioned above, the uncertainties in the orbital elements grow large due to the effect of close encounters with Venus. This is confirmed by the number of close approaches with Venus recorded by the \texttt{mercury} integrator. The left panel of Fig.~\ref{fig:Venus_close_encounters} shows all the close approaches recorded within 0.02 au as a function of time. Clusters around $-11$ kyr for 2025~GN1 and around $-12$ kyr for 2021~PH27 can be clearly seen, with a small fraction of encounters reaching distances smaller than 0.001 au. Note that encounters closer than about 0.004 au are well separated in time. 

To analyse possible separation events, we followed the procedure outlined in \citet{Pravec2019}. It consists in computing the distances and relative velocities between all the possible couples of clones of the two asteroids, and then checking for occurrences of minimum distances with an order of magnitude comparable to the Hill radius of the largest component.
Note that the Hill radius \citep{Hamilton1992} of 2021~PH27 can be estimated to be roughly between 15 km and 30 km.
By using the one-day timestep output obtained from the numerical integrations, we computed the distance and relative velocity for all the pairs of simulated clones, and we then recorded all the encounters closer than 200~000 km. The scatter plot in the plane of year and close encounter distance is shown in the right panel of Fig.~\ref{fig:Venus_close_encounters}. The colour of the marker indicates the close encounter relative velocity. Many close encounters between clone pairs are found between today and $-10$ kyr at a relative velocity of about 1 km/s; however, they are all at distances larger than 150~000 km, which is about 10000 times the Hill radius of 2021~PH27. After 10 kyr of backwards evolution, close encounters still appear even at smaller distances; however, they are randomized as a result of close encounters with Venus seen above, causing a large dispersion in the distribution of the close encounters. Therefore, we conclude with near certainty that the pair separated more than 10.5 kyr ago. Close approaches within 10-100 times the Hill radius of 2021~PH27 occur between $-10$ kyr and $-100$ kyr, however they are not statistically significant for identifying a precise separation age. 

\begin{figure}
    \centering
    \includegraphics[width=0.98\textwidth]{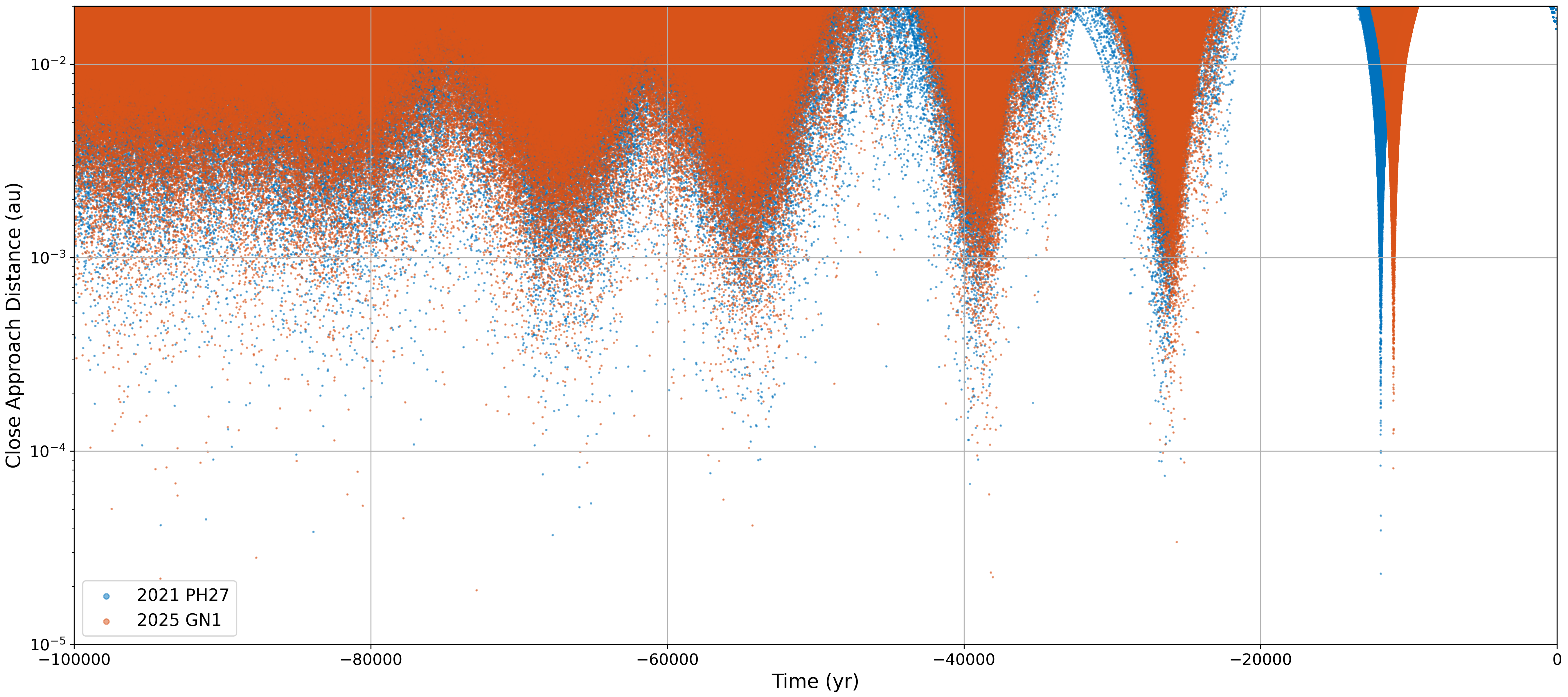}
    \includegraphics[width=0.98\linewidth]{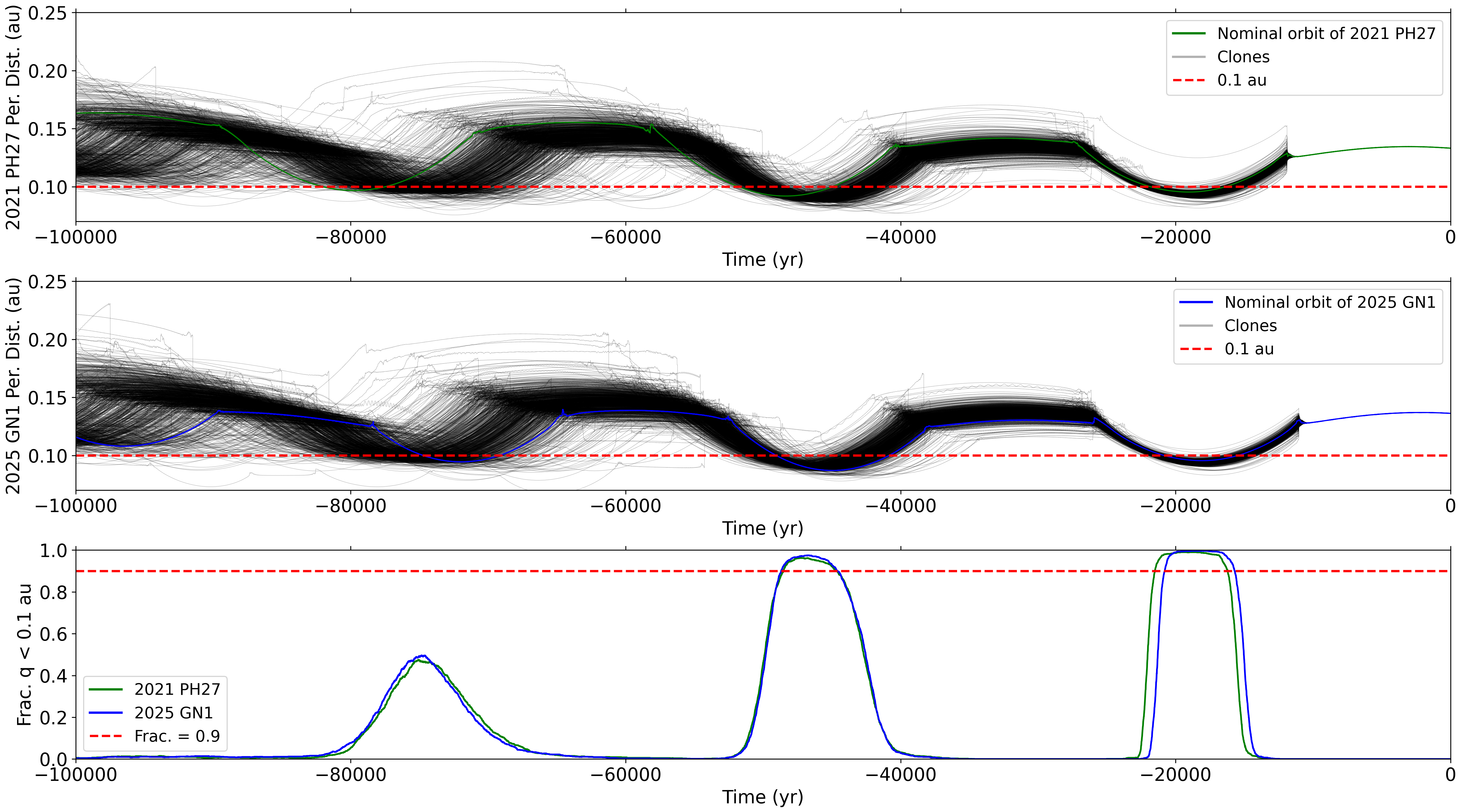}
    \caption{On the top panel, distance of the recorded close encounters with Venus, as a function of time, for the extended backwards integrations over 100 kyr. Blue dots are close encounters registered for 2021~PH27, while the red dots are those for 2025~GN1. On the bottom panel, the perihelion distance of 2021~PH27 (top subpanel) and of 2025~GN1 (middle subpanel). The nominal orbit of 2021~PH27 is reported in green, while that of 2025~GN1 in blue. Clones are represented by the gray curves. In these two subpanels, the red dashed line indicates a perihelion distance of 0.1 au. The bottom subpanel represents the fraction of clones reaching a perihelion distance smaller than 0.1 au as a function of time. The red dashed line here represent a fraction of 0.9.}
    \label{fig:perihelion_long}
\end{figure}

To better understand the past dynamical evolution of the two asteroids before the last Venus close approach, we extended the numerical integrations back to $-100$ kyr, using the same time step of 0.5 days, but recording the output orbital elements only every 10 years. Figure~\ref{fig:perihelion_long}, top panel, shows the distribution of Venus' close approaches. While they undergo a complex history of close approaches with Venus, we can see clear patterns in the distribution of close approaches in terms of time and distance. Five different close approach times can be clearly seen, with the last one at about $-65$ kyr. After this time of backwards evolution, the dynamics randomize almost completely, and patterns in close approaches with Venus start to be fuzzy as well.

Another interesting quantity to keep track of during the backwards evolution is the perihelion distance $q$, because it could be connected to a possible disruption mechanism at low perihelion distance \citep{Granvik2016}. The bottom panel of Fig.~\ref{fig:perihelion_long} shows the evolution of the perihelion distance of the 1000 clones of each asteroid, together with their nominal evolution. A red dashed horizontal line is plotted at perihelion distance of 0.1 au, which corresponds to the catastrophic disruption limit of asteroids of similar size to 2025~GN1 \citep{Granvik2016}. Also, the percentage of the number of clones reaching a perihelion distance smaller than 0.1 au, as a function of time, is reported in the bottom panel. Despite the complex history of Venus' close approaches, there are two clear instances at which $q < 0.1$ au with a probability larger than 90\%, with some instances reaching peaks $>99$\%. The first one happened between about $-17$ and $-21$ kyr, while the second one happened between $-45$ and $-48$ kyr. After this time, chaotic effects grow large and the dynamics of the clones follow less closely the secular evolution \citep{gronchi-milani_2001, fenucci-etal_2022}, and large dispersions are seen in the orbital elements.

\section{Discussion on possible separation mechanisms}
\label{sec:discussion}
As we have seen in the previous sections, 2021 PH27 and 2025 GN1 have very similar Keplerian elements for at least 40-100 kyr in the past and are of identical spectral type, and these support a common origin as also found by \cite{Sheppard2025}. Barring the improbable event of a collision with another asteroid, there are at least four possible disaggregation mechanisms that could affect an asteroid with a low perihelion and subject to close passes with the planet Venus:

\begin{enumerate}
\item Venus flybys below the Roche limit, resulting in tidal fragmentation.
\item Tidal fragmentation driven by the Sun due to low perihelion value.
\item Fragmentation following the increase in surface temperature due to the low perihelion.
\item The ex binary: YORP effect, which increases the asteroid's rotation frequency up to the spin barrier value, forms a binary system which subsequently separates into two components.
\end{enumerate}

\noindent We now discuss whether these fragmentation mechanisms can explain the formation of the asteroid pair 2021~PH27-2025~GN1. 

\subsection{Tidal fragmentation at Venus flybys}
The numerical integrations of the orbital elements (see Fig.~\ref{fig:clo_elements} and Fig.~\ref{fig:perihelion_long}) show that the orbits of the two asteroids remain similar until the last close encounter with Venus of 2025 GN1 occurred about $-$11 kyr. Previous to this epoch, the nominal orbits diverge and the clouds of clones spread in the space of the Keplerian elements. If we see the close approaches with Venus of 2021 PH27 at $-12$ kyr (Fig.~\ref{fig:Venus_close_encounters} on the left), it can be seen that most of the clones have a distance equal to or greater than 0.001 au, about 150~000 km. A tidal disruption is determined by the slightest distance $d$ and by the relative speed $v_\infty$ between the asteroid and the planet. The lower these values, the greater the probability of fragmentation because the flyby time increases. Venus contributes about 1/3 to the fragmentation of the NEAs; the major role is played by the Earth because, at Venus' distance, the NEAs are faster and the flyby time is shorter \citep{Schunova2014}.\\
The Roche limit of the planet $d_{Roche}$ is the distance from the center of the planet below which a cohesionless rubble-pile asteroid cannot maintain its structure and is broken apart into its component blocks \citep{Roche1849, chandrasekhar1969}: 

\begin{equation}
    d_{Roche}=2.46R_{p}\sqrt[3]{\frac{\rho_p}{\rho}}
    \label{eq:roche}
\end{equation}

\noindent In Eq.~\eqref{eq:roche}, $R_p=6052~\text{km}$ and $\rho_p=5.24~\text{g}/\text{cm}^3$ are the radius and the mean density of the planet (Venus in this case), while $\rho$ is the mean density of the asteroid making the flyby. Therefore, for a rubble-pile asteroid with average density between 1 and $3 ~\text{g}/\text{cm}^3$, $d_{Roche}$ ranges from 26,000 to 18,000 km from the center of Venus. 
The minimum distance value for Venus close approaches happening at the same epoch for 2021~PH27 and 2025~GN1 can be obtained by superimposing the two close approaches shown in the left panel of Fig.~\ref{fig:Venus_close_encounters}. This corresponds to about $0.004$ au, which converts to $\sim$600,000 km. This value is about 20 times larger than the Roche limit of Venus, so we can consider very unlikely that the pair was born as a result of a past Venus flyby. If we go even further back in time up to 100 kyr (Fig.~\ref{fig:perihelion_long} top panel), we can see that there are at least four other flybys with Venus, without, however, the close approach distance dropping significantly below 0.001 au. 

\subsection{Sun tidal fragmentation}
The same logic ad the previous section can be applied to tidal fragmentation for passages close to the Sum. Both asteroids experience perihelion distances between 0.1 and 0.08 au, occurring in the time intervals $(-17, -21)$ kyr and $(-45,-48)$ kyr, see Fig.~\ref{fig:perihelion_long} bottom panel. The Roche limit of the Sun for an asteroid with the same density range as considered above is 1.3 to 2 million km, well below the minimum value of about 0.08 au (12 million km). Therefore, tidal fragmentation due to the Sun can be also considered unlikely as with tidal fragmentation by Venus close encounters, thus eliminating the first two fragmentation mechanisms. We next consider the third mechanism.

\subsection{High temperature fragmentation}
Although the perihelion values are well above the Roche limit of the Sun, they are still in the range where super-catastrophic breakup occurs \citep{Granvik2016}. NEAs with low perihelion values are fewer in number than expected based on arrivals from the main belt: population models predict a greater number of asteroids with perihelion close to the Sun than are actually observed. The observed number of NEAs is reconciled with the predictions if we assume that, below a certain critical perihelion distance $q_*=0.076 \pm 0.0025$ au, asteroids are subject to catastrophic destruction. Analyzing observed asteroids with perihelion $q$ between 0.2 au and 0.02 au, \cite{Granvik2016} found that there is an inverse correlation between the $q_*$ value and the asteroid's diameter: larger asteroids have $q_*$ closer to the Sun than smaller ones, which have higher $q$ values. This indicates that fragmentation occurs at greater distances for small asteroids, while larger ones fragment at smaller distances. \\
The mechanism that can destroy an asteroid with a perihelion too close to the Sun, but above the Roche limit, is not yet clear, but it appears to be related to the temperature its surface can reach, and there can be several mechanisms at work simultaneously. It could be a consequence of thermal fracturing of the surface: fragments are swept away by radiation pressure, and the asteroid loses mass orbit after orbit. Or it could depend on the anisotropic emission of thermal photons, i.e. the YORP effect, or the anisotropic emission of gas molecules that sublimate and increase the rotation frequency until they break up, mimicking the YORP effect. Alternatively, asteroids can contain volatile materials that, during sublimation, exert sufficient force to fragment the body. The systematic destruction of asteroids with low perihelions explains why the Atens have, on average, higher albedos than the Apollos and the Amors. Atens with lower albedos belong to more primitive taxonomic classes, such as the C-type, which have an amount of volatiles inside them \citep{Trigo2019}, and these, during the sublimation phase, can exert a separation pressure between the asteroid's constituent blocks. In fact, C-type asteroids may be richer in water than previously thought: isotopic analyses of rock samples taken by Hayabusa 2 from the surface of the C-type asteroid (162173) Ryugu indicate that the asteroid's rocks have been subjected to liquid water flow \citep{Lizuka2025}. Ryugu is likely an extinct comet \citep{Miura2022}, and this discovery would at least double the amount of water brought to Earth by carbonaceous asteroid impacts \citep{Lizuka2025}. Alternatively, because low albedo asteroids are darker, surface thermal fracturing is more efficient, and the asteroid disintegrates rapidly.\\
Given that 2021 PH27 and 2025 GN1 had perihelion between 0.1 and 0.08 au, which occurred at $-17$/$-21$ kyr and $-45$/$-48$ kyr, it is possible that the parent asteroid was broken apart because it was located within a region dangerous for the integrity of asteroids. After the formation of the asteroid pair, the emission of dust particles and small meteoroids from the surface probably continued due at least to thermal fracturing. Laboratory experiments with the Space and High-Irradiance Near-Sun Simulator (SHINeS) \citep{Tsirvoulis2022} have shown that subjecting CI-type chondrites, which compose the surface of many C-type asteroids such as Ryugu \citep{Yokoyama2023}, to the same irradiation conditions found between 0.06 and 0.23 au from the Sun results in spontaneous fracturing \citep{Schirner2024} at velocities that can exceed the low escape velocity of a small asteroid, which would thus release meteoroids into space\footnote{\url{https://meetingorganizer.copernicus.org/EPSC-DPS2025/EPSC-DPS2025-554.html}}. As a comparison, we can cite (3200) Phaethon, whose emission of micrometre-sized dust particles has been repeatedly observed during its perihelion passage. Although it is not clear which mechanism is dominant, thermal fragmentation, sodium emission, or electrostatic ejection, it is undeniable that this is related to the low perihelion value \citep{Kimura2022, Man-To2024}. So associated with the 2021~PH27-2025~GN1 pair, it is plausible that there is also a stream of meteoroids sharing their orbit, which is supplied with material at each perihelion passage.\\
In this scenario, we might expect to see areas on the surfaces of the two asteroids that have undergone less space weathering than the rest. However, to highlight these areas, the entire surface would need to be spectroscopically mapped during an entire rotation to determine whether spectral changes are occurring that cause surface rejuvenation. For example, in the case of S-type asteroids, surface rejuvenation causes them to transition into Q-type asteroids \citep{DeMeo2023}.

\subsection{The ex binary hypothesis}
\begin{figure}
    \centering
    \includegraphics[width=1.00\textwidth]{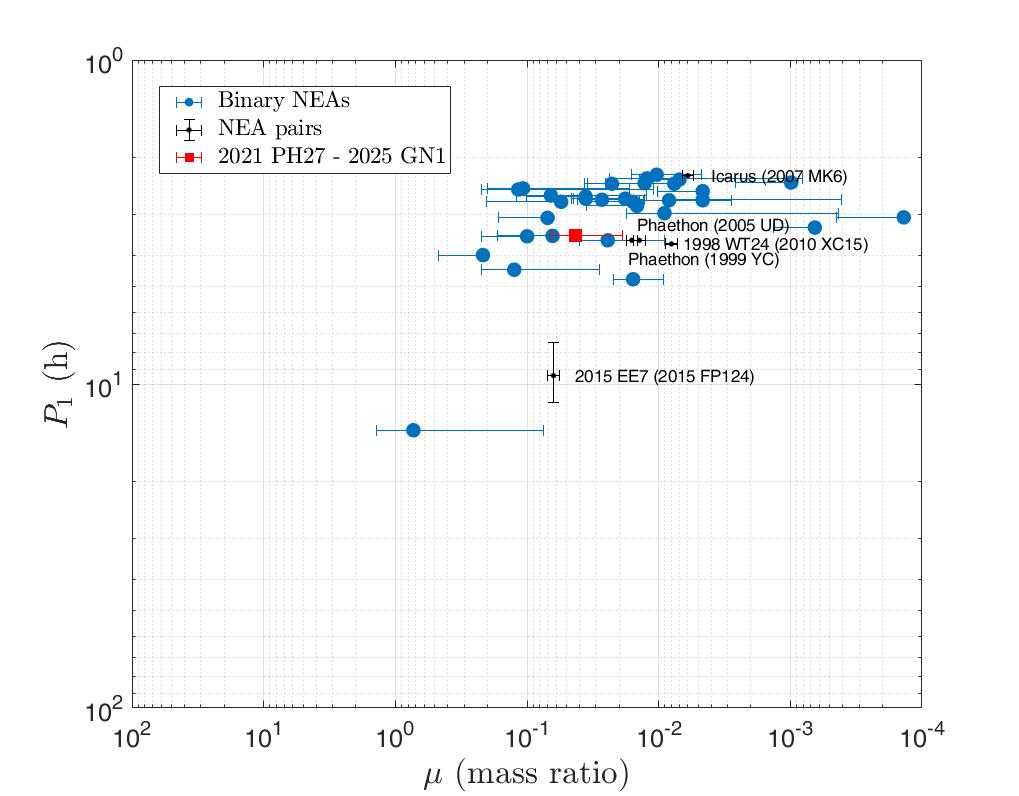}
    \caption{Plot of the primary rotation periods as a function of the mass ratio for the best-known 32 NEA binary system. The uncertainties in the rotation period of the primary are not perceptible at this scale, while the uncertainty in the mass ratio is a consequence of the uncertainty in the diameters of the primary and the satellite. The square is the pair 2021~PH27-2025~GN1, while the black dot are the pairs of Fig.~\ref{fig:NEA_pairs_plot}. The asteroid 2015~EE7 has an uncertain rotation period. The mass ratio value of 0.2 separates the region of stable binary systems (left) from unstable ones (right), which will become asteroid pairs.}
    \label{fig:pairs_plot}
\end{figure}

Now let's go deeper into the possibility that the pair 2021 PH27-2025 GN1 originated from the destruction of a binary system formed by the rotational disintegration of a parent asteroid by the YORP effect or gas emission. To test whether the physical characteristics of the two asteroids are consistent with an ex-binary system, we plotted the primary period versus mass ratio for all 32 known binary NEAs with reasonably well-determined pair parameters, including the rotation period of the primary and the diameter of the two components with the respective uncertainty. The data were taken from the archive maintained by \cite{Johnston2019}, updated to 18 Dec 2025 \footnote{\url{https://www.johnstonsarchive.net/astro/asteroidmoons.html}}. As a rotation period of the primary, i.e., 2021~PH27, we take $3.49 \pm 0.01$ h, while the mass ratio $\mu = 0.04 \pm 0.02$ is given by Eq.~\eqref{eq:mass_ratio}, in the hypothesis that the average density is the same. The results are shown in Fig.~\ref{fig:pairs_plot}. The red square representing 2021~PH27-2025~GN1 overlaps well with the general trend of the binary NEAs and the other NEA pairs; this supports the hypothesis that 2021~PH27 was originally a binary system and that the pair was born by rotational dissociation \citep{Pravec2010}. With a rotation period of $3.49 \pm 0.01$ h, from Eq.~\eqref{eq:limit_period} in the hypothesis that the asteroid is near the cohesionless spin-barrier, we can obtain a rough estimate of the mean density of 2021~PH27 that results in $ 0.895 \pm 0.005~\text {g}/\text{cm}^3$, a value similar to (101955) Bennu, a classical rubble-pile asteroid \citep{Lauretta2019}.

Due to the chaotic dynamics induced by close encounters with Venus, we are unable to give a value for the separation age. We can only say, with certainty, that it occurred at least 10.5 kyr ago. However, we can speculate that it happened during the last low perihelion phase of $17$/$21$ kyr ago. This phase could have led to an increase in the YORP effect during Sun proximity, and it is possible that surface warming also played a role. This recent separation would explain why the orbital elements of the two asteroids are still so close together. In fact, if the fragmentation had occurred during the previous low perihelion period, between 45 and 48 kyr ago, the orbital elements today would be statistically more dispersed due to three Venus close encounters (see Fig.~\ref{fig:perihelion_long}, top panel), which would be in contradiction with today's orbital elements. Finally, the age of $10$ kyr estimated by \cite{Sheppard2025} is not supported by our results.\\
This recent fragmentation event, if it occurred, likely dispersed a large number of small fragments into the orbit of this asteroid pair. Considering a fragment of about 0.1 cm in diameter with a mean density of 3.5 g/cm$^3$, we can estimate a Poynting-Robertson decay rate in the semimajor axis $a$ with the equation \citep{Wyatt1950}:

\begin{equation}
    \frac{da}{dt}=-\frac{\alpha\left(2+3e^2\right)}{a\left(1-e^2\right)^{3/2}}.
    \label{eq:poynting_robertson}
\end{equation}

\noindent In Eq.~\eqref{eq:poynting_robertson} $e\approx 0.71$ is the orbital eccentricity while $\alpha\approx 2\cdot 10^{-7}$ au$^2$/yr is a constant for the assumed meteoroid. With these data we have $da/dt\approx -4.4 \cdot 10^{-6}$ au/yr, so the expected stream lifetime with $a\approx 0.46$ au is about  $t\approx a/(da/dt)\approx 100$ kyr. If the orbit were supposed to be circular, the fall time would be about 3 times faster. With this rough estimate, we can say that it is possible that the pair was still associated with the older meteoroid stream. At this older stream, new meteoroid emissions can be superimposed due to its low perihelion passage near the Sun, so we can reasonably expect to have a meteor shower in Venus' atmosphere associated with 2021 PH27-2025 GN1.

\subsubsection{Time scale of the YORP effect}
It is not yet clear on what time scale the YORP effect was able to modify the rotation period of the parent asteroid of the pair 2021~PH27-2025~GN1. To estimate this, we calibrated a scaling relation to get the order of magnitude of the time required for the rotation period of the parent asteroid to double. As a parent body, given that the ex-satellite 2025~GN1 has a significantly lower mass than 2021~PH27, we assume a body similar to 2021~PH27, i.e., with the same dimensions and mean density. The mean change in angular rate due to the YORP effect is given by \citep{Carbognani2011}:

\begin{equation}
    \frac{d\omega}{dt}\approx C\frac{\left(1-A\right)}{\rho a^2 D^2}
    \label{eq:yorp}
\end{equation}

\noindent In Eq.~\eqref{eq:yorp} $C$ is a constant that depends on the asteroid's geometry and mass distribution, $A$ is the Bond albedo, $\rho$ is the mean asteroid density, $a$ the semimajor-axis and $D$ the asteroid's effective diameter. To get the order of magnitude of $C$, we used the measured value of $d\omega/dt$ due to the YORP effect for some asteroids: (1862) Apollo, (161989) Cacus and (1685) Taurus \citep{Nesvorny2008, Durech2018}. Apollo is a Q-type asteroid with a diameter of about 1500 m, Cacus is an S-type with a diameter of about 1900 m, and finally Toro is still an S-type asteroid 3.4 km wide. \\
The $C$ values found in these three cases are close to each other, with an average of $C\approx 2\pm 0.6\cdot 10^{22}$ kg m $\text{s}^{-1}$ $\text{yr}^{-1}$. With this approximate value, assuming that the geometry and mass distribution of 2021 PH27 is not too different from the asteroids considered as reference, and taking $a=0.46$ au (as can be seen from Fig.~\ref{fig:clo_elements} the semi-major axis remains relatively constant on time scales of the order of 100 kyr), $\rho\approx 1000$ kg $\text{m}^{-3}$, $D\approx 1.2$ km and $A\approx 0.04$, we get $d\omega/dt\approx 3\pm 1\cdot 10^9$ $\text{s}^{-1}$ $\text{yr}^{-1}$, so the doubling of the actual rotation period is possible in about $150 \pm 50$ kyr. It is reasonable to think that the YORP effect was actually able to increase the rotation frequency enough to fragment the parent asteroid during the parent's dynamical lifetime.\\

\section{Conclusions}
\label{sec:end}
In this paper, we investigated the origin of the NEA pair 2021~PH27-2025~GN1. Based on the available orbital elements, its dynamical evolution in the past has been reconstructed up to -100 kyr, finding a similar trend for the orbital elements. In the last 100 kyr, there have been five flybys with Venus, but at minimum distances so as not to have tidal disruption if the average mean density is in the range 1-3 g/cm$^3$. Even the perihelion values are not low enough to enter the Roche limit of the Sun. \\
Instead, both asteroids have undergone a similar orbital evolution, and the perihelion has dropped in the range 0.1-0.08 au, which occurred at $-17$/$-21$ kyr and $-45$/$-48$ kyr. This leads to enhanced thermal fragmentation, gas and thermal photon emission from the surface. The rotation period of the primary and the mass ratio it is indistinguishable from known binary systems among NEAs. This fact suggests an origin from rotational disintegration due to the YORP effect or gas emission of a rubble pile during low perihelion phases. This disintegration formed an unstable binary system with a low mass secondary that, separating, formed the pair of asteroids that we see today. \\
Due to close encounters with Venus that make their orbits chaotic, it has not been possible to establish an age of separation; it can only be said that the pair is older than -10.5 kyr, but since the low perihelion phase between -17/-21 kyr has a higher probability of having perihelion less than 0.1 au, it is likely that it occurred in this time interval. This recent fragmentation would explain why the orbital elements are still so similar. The low perihelion value makes it plausible that meteoroids could be emitted from the surface, as happens at (3200) Phaethon, so the existence of a meteoroid stream, also due to the possible recent fragmentation and associated with the pair, is very likely. If it exists, this meteoroid stream causes, due to the small value of the MOID, a meteor shower in the atmosphere of Venus.

\section*{Acknowledgements}
This research has made use of data and/or services provided by the International Astronomical Union's Minor Planet Center. TSR acknowledges funding from Ministerio de Ciencia e Innovación (Spanish Government), PGC2021, PID2021-125883NB-C21. This work was partially supported by the Spanish MICIN/AEI/10.13039/501100011033 and by "ERDF A way of making Europe" by the “European Union” through grant PID2021-122842OB-C21, and the Institute of Cosmos Sciences University of Barcelona (ICCUB, Unidad de Excelencia ’María de Maeztu’) through grant CEX2019-000918-M.
Based on observations made at the international Gemini Observatory, a program of NSF NOIRLab, is managed by the Association of Universities for Research in Astronomy (AURA) under a cooperative agreement with the U.S. National Science Foundation on behalf of the Gemini partnership: the U.S. National Science Foundation (United States), the National Research Council (Canada), Agencia Nacional de Investigación y Desarrollo (Chile), Ministerio de Ciencia, Tecnología e Innovación (Argentina), Ministério da Ciência, Tecnologia, Inovações e Comunicações (Brazil), and Korea Astronomy and Space Science Institute (Republic of Korea).
Gemini GS-2025A-DD-108 data was acquired through the Gemini Observatory Archive \citep{2017ASPC..512...53H} at NSF NOIRLab and processed using DRAGONS (Data Reduction for Astronomy from Gemini Observatory North and South; \citet{2023RNAAS...7..214L} and \href{https://zenodo.org/records/10841622}{Zenodo}).

\section*{Data Availability}
The data underlying this paper will be shared on reasonable request to the corresponding author.



\bibliographystyle{cas-model2-names}
\bibliography{Dynamicbiblio}{} 

%%%%%%%%%%%%%%%%%%%%%%%%%%%%%%%%%%%%%%%%%%%%%%%%%%

\end{document}